\documentclass[final,3p,times,twocolumn]{elsarticle}

\usepackage{epsfig}

\biboptions{}

\journal{Physics Letters B}

\begin{document}

\begin{frontmatter}



\title{Associated heavy quarks pair production with Higgs  as a tool for
a search for non-perturbative effects of the electroweak interaction
at the LHC}
\author[a]{B.A. Arbuzov}
\ead{arbuzov@theory.sinp.msu.ru}
\author[a]{I.V. Zaitsev}
\address[a]{M.V. Lomonosov Moscow State University,
 119991 Moscow, Russia}



\begin{abstract}
Assuming an existence of the anomalous triple electro-weak bosons
interaction being defined by coupling constant $\lambda$ we calculate
its contribution to interactions of the Higgs with pairs of heavy
particles. Bearing in mind experimental restrictions $-0.011 <
\lambda < 0.011$ we present results for possible effects in processes
$p\,p \to W^+ W^- H,\,p\,p \to W^+ Z H,\,p\,p \to W^- Z H,\,p\,p \to
\bar t t H $, $p p \to \bar b b H$. Effects could be significant with
negative sign of $\lambda$ in associated heavy quarks $t,\,b$ pairs
production with the Higgs. In calculations we rely on results of the
non-perturbative approach to a spontaneous generation of effective
interactions, which defines the form-factor of the three-boson
anomalous interaction.
\end{abstract}


\end{frontmatter}

\section{Introduction}\label{sec1}

The totality of data nowadays confirms main features of the Standard
Model, which consists of QCD, describing strong interactions, and the
EW theory, describing electroweak interactions. This confirmation is
essentially based on numerous perturbative calculations which
describe corresponding data. However in QCD the inevitable
introduction of non-perturbative effects is also evident. First of
all the low momenta region of the strong interaction definitely can
not be described in the framework of the perturbative calculations.
Examples of non-perturbative quantities are well-known: vacuum
averages the gluon condensate $<\frac{\alpha_s}{\pi}G_{\mu
\nu}^aG_{\mu \nu}^a>$, the quark condensate $< \bar q\,q >$ {\it
etc}. One of the most powerful methods of dealing with the
non-perturbative effects is provided in the framework of approaches
using the so-called effective interactions. The eldest and the most
popular such effective interaction is the famous Nambu--Jona-Lasinio
interaction \cite{NJL,NJL2}. With application to quark structure of
hadrons this approach adequately describes the low momenta region,
see {\it e.g.} reviews \cite{Volb, ERV, VolRad}.

The Nambu--Jona-Lasinio interaction deals with four-quark effective
terms. However, the non-zero gluon condensate testifies for
additional effective terms also in gluon interactions. There were
also proposals for such terms. In particular, triple gluon
interaction in the low momenta region of the following form
\begin{equation}
L_{eff}\,=\,-\frac{G}{3!}f_{abc}G^a_{\mu\nu}G^b_{\nu\rho}G^c_{\rho\mu};
\label{3gluon}
\end{equation}
where $G^a_{\mu\nu}$ is a gauge covariant gluon field and $f_{abc}$
are structure constants of the color $SU(3)$, was proposed in
work~\cite{aab}.

In the electro-weak theory necessity of non-perturbative contribution
is not nowadays so evident as in QCD. However the structure of gauge
theories is the same for both cases. One might expect similar
features in three-boson interactions. In particular, the following
three weak boson interaction was introduced ~\cite{Hag1, Hag2}
\begin{eqnarray}
& &L_{eff}\,=\,-\frac{G_W}{3!}F(p_i)\epsilon_{abc}W^a_{\mu\nu}W^b_{\nu\rho}W^c_{\rho\mu};
\label{3boson}\\
& &G_W=\frac{g\, \lambda}{M_W^2};\; W_{\mu \nu}^3= \cos\theta_W Z_{\mu\nu} +\sin\theta_W\,A_{\mu\nu};\nonumber
\end{eqnarray}
where $g$ is the electro-weak gauge coupling and indices $a,b,c$ take
three values and the third boson $W_{\mu \nu}^3$ is a composition of
neutral bosons $Z$ and $\gamma$. Form-factor $F(p_i)$ in~\cite{Hag1, Hag2} is postulated
and it has to vanish for $|p^2_i|\gg\Lambda^2$, where $\Lambda^2$ is a
characteristic scale. Interaction~(\ref{3boson}) would
lead to effects {\it e.g.} in electro-weak bosons pair production and
was studied in experiments. The best limitations for parameter
$\lambda$ is provided by recent data of CMS
Collaboration~\cite{CMSlam}
\begin{equation}
-0.011<\lambda<0.011.\label{CMSlam}
\end{equation}
Both the Nambu--Jona-Lasinio interaction and
interaction~(\ref{3gluon}) are supposed to act in a low momenta
region. This means, that in both cases form-factors are present,
which guarantee decreasing of intensity of the interactions for large
momenta. In the original NJL \cite{NJL,NJL2} interaction a cut-off
was introduced for the purpose. Starting of fundamental gauge
theories of interactions of the Standard Model we have to understand
the origin of such effective cut-off. This can be done under
assumption of these interactions being spontaneously generated. The
notion of spontaneous generation is traced back to methods of the
superconductivity theory. In application to superconductivity the
conception of compensation principle was elaborated \cite{Bog1, Bog2}
by N.N. Bogoliubov. This approach was applied to spontaneous
generation of Nambu--Jona-Lasinio interaction in work \cite{AVZ06}
and of interaction~(\ref{3gluon}) in work \cite{AZ20132}. In the
course of this application form-factors inherent to corresponding
interactions are uniquely defined \footnote{Of course, in the
framework of an approximation.}. As an additional confirmation of
applicability of the method to non-perturbative quantities, value
$V_2$ of the gluon condensate was calculated \cite{AZ20132} in
agreement with its phenomenological value.
\section{Additional interactions of the Higgs with electro-weak bosons}
\label{sec2}

\begin{figure}
\includegraphics[width=8.5cm]{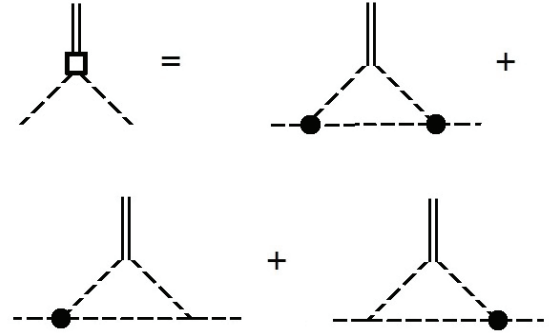}
\caption{Diagram representation of non-perturbative contribution
to $V V H$ vertices depicted as square.
Double lines represent Higgs boson,
dotted lines represent electro-weak bosons. Black spots represent
non-perturbative vertex~(\ref{3boson}, \ref{VW}) and simple points corresponds
to usual SM interaction.}
\label{fig:HWW}
\end{figure}

Electro-weak bosons $W^\pm,\,Z$ due to their large masses interact
with Higgs $H$ significantly. Namely there are the following vertices
for interaction of Higgs $H$ with $W^+W^-$ and $Z\,Z$ respectively
\begin{eqnarray}
& &\imath\,g M_W\,g_{\mu\nu};\quad \imath \frac{g M_Z}{\cos\theta_W}\,
g_{\mu\nu}.\label{V0}
\end{eqnarray}
Now let us assume, that in addition to usual three-boson gauge
interaction, interaction~(\ref{3boson}) really exists. Then
three-boson vertex takes the form

\begin{eqnarray}
& &V(\mu,p;\nu,q;\rho,k)= - g\,\epsilon_{abc}\bigl(g_{\mu\nu}
(q_{\rho}-p_{\rho})+\nonumber\\
& &g_{\nu\rho}
(k_{\mu}-q_{\mu})+g_{\rho\mu}(p_{\nu}-k_{\nu})+
\frac{\lambda}{M_W^2} F(p,q,k)\times\nonumber\\
& &(g_{\mu\nu}(q_{\rho} pk-p_{\rho}\, qk)+g_{\nu\rho}
(k_{\mu} pq -q_{\mu} pk)+\label{VW}\\
& &g_{\rho\mu}(p_{\nu} qk -k_{\nu} pq )+
q_{\mu}k_{\nu}p_{\rho}-k_{\mu}p_{\nu}q_{\rho})\bigr).\nonumber
\end{eqnarray}
Here $F(p,q,k)$ is a form-factor, which is defined in the framework
of the spontaneous generation of effective
interaction~(\ref{3boson})~\cite{AVZ2} in the compensation approach,
which we have discussed in the Introduction. Then we have additional
contribution to $V V H$ vertex due to terms proportional to $\lambda$
and to $\lambda^2$ according to diagrams presented in
Fig.~\ref{fig:HWW}. As a consequence we have additional contribution
to vertices $V_{VV'H}$, where $V$ and $V'$ correspond to
electro-weak bosons $W, Z, \gamma$.

The combined account of both vertices~(\ref{V0}, \ref{VW}) leads to
corrections for $V V' H$ vertex according to diagrams shown in
Fig.~\ref{fig:HWW}.
We have for vertices of interactions of the Higgs with two
electro-weak bosons $V(p,\mu)\,V'(q,\nu)$ instead of~(\ref{V0})
\begin{eqnarray}
& &V_{W^+W^-H}=\imath g M_W(g_{\mu\nu}+G_{WW}
(g_{\mu\nu}pq-p_{\nu}q_{\mu}));
\nonumber\\
& &V_{ZZH}=\imath\frac{g M_Z}{\cos\theta_W}(g_{\mu\nu}+G_{ZZ}
(g_{\mu\nu}pq-p_{\nu}q_{\mu}));
\nonumber\\
& &V_{ZAH}=\imath\frac{g M_Z}{\cos\theta_W}G_{Z\gamma}\,(g_{\mu\nu}pq-p_{\nu}q_{\mu});
\label{VVH}\\
& &V_{AAH}=\imath\frac{g M_Z}{\cos\theta_W}G_{\gamma\gamma}\,(g_{\mu\nu}pq-p_{\nu}q_{\mu});
\nonumber\\
& &G_{Z\,Z}=\cos^2\theta_W\,G_{W_0 W_0};\nonumber\\
& &G_{\gamma\,\gamma}=\sin^2\theta_W\,G_{W_0 W_0};\nonumber\\
& &
G_{Z\,\gamma}=2\cos\theta_W \sin\theta_W\,G_{W_0 W_0};\nonumber\\
& &G_{W_0 W_0}=\frac{2 g}{M_W^2}\Bigl(\frac{3 g \lambda}{8 \pi^2}I_1-
\frac{\sqrt{2 \lambda^2}}{\pi}I_2\Bigr);\nonumber\\
& &G_{W\, W}=\frac{g}{M_W^2}\Bigl(\frac{3 g \lambda}{8 \pi^2}I_{W1}-
\frac{\sqrt{2 \lambda^2}}{\pi}I_{W2}+\nonumber\\
& &\frac{3 g \lambda}{8 \pi^2}I_{Z1}-
\frac{\sqrt{2 \lambda^2}}{\pi}I_{Z2}\Bigr).\nonumber
\end{eqnarray}
Here
\begin{eqnarray}
& &I_1 = \int^{z_0}_0\frac{F(t)\, dt}{2(\sqrt{t}+\mu)^2};
\, I_2 = \int^{z_0}_0\frac{2 t F^2(t)\,dt}{(\sqrt{t}+\mu)^3};\nonumber\\
& &I_{W1} = \int^{z_0}_0\frac{F(t)(\sqrt{t}+s\mu_Z)\,dt}
{(\sqrt{t}+\mu_Z)^2(\sqrt{t}+\mu)};\nonumber\\
& &I_{W2} = \int^{z_0}_0\frac{F^2(t)\sqrt{t}\,(\sqrt{t}+s\mu_Z)\,dt}
{(\sqrt{t}+\mu_Z)^2(\sqrt{t}+\mu)};\label{int}\\
& &I_{Z1} = \int^{z_0}_0\frac{F(t)\sqrt{t}\,dt}
{(\sqrt{t}+\mu_Z)^2(\sqrt{t}+\mu)(1-s)};\nonumber\\
& &I_{Z2} = \int^{z_0}_0\frac{F^2(t)\,t\, dt}
{(\sqrt{t}+\mu_Z)^2(\sqrt{t}+\mu)(1-s)};\nonumber\\
& &\mu=\frac{g |\lambda|}{16 \sqrt{2} \pi};\quad \mu_Z=\frac{g |\lambda|}
{16 \sqrt{2} \pi\, (1-s)};\nonumber\\
& &t=\frac{G_W^2 (p^2)^2}{512 \pi^2};\;s = \sin^2 \theta_W.\nonumber
\end{eqnarray}

We take form-factor $F(t)=F(p,-p,0)$ from results of work~\cite{AVZ2}
in which the compensation approach was applied to the electro-weak
interaction:
\begin{eqnarray}
& &F(t)=\frac{1}{2}G^{31}_{15}\bigl(\,t|^{\,0}_{1,1/2,0,-1/2,-1}\bigr)-
\frac{85\sqrt{2}g_0}{128 \pi}\times\nonumber\\
& &G^{31}_{15}\bigl(\,t|^{1/2}_{1,1/2,1/2,-1/2,-1}\bigr)+
C_1 G^{10}_{04}\bigl(\,t|\frac{1}{2},1,-\frac{1}{2},-1\bigr)+\nonumber\\
& &C_2 G^{10}_{04}\bigl(\,t|1,\frac{1}{2},-\frac{1}{2},-1\bigr);\;
t=\frac{G_W^2 \,(p^2)^2}{512\,\pi^2};\label{FF}\\
& &F(t)= 0,\, t\geq t_0;\quad t_0=9.6175,\nonumber\\
& &g_0=0.6037,\,C_1=-0.0351,\,C_2=-0.0511.\nonumber
\end{eqnarray}
Here $g_0$ is the value of gauge electro-weak coupling $g$ at point
$t=t_0$ and  we use Meijer functions
$$G^{mn}_{pq}\bigl(\,t|\,^{a_1,...a_p}_{b_1...,b_q}\bigr)\,,$$ for more
details see, {\it e.g.} book~\cite{ABOOK}. Characteristic scale $\Lambda$,
corresponding to form-factor~(\ref{FF}) is defined by the following expression
\begin{equation}
\Lambda^4\,=\,\frac{512 \pi^2\,t_0}{g_0^2\,\lambda^2}\,M_W^4\,;\label{Lambda}
\end{equation}
and {\it e.g.} $\Lambda\,=\,19.83\,TeV$ for $|\lambda|=0.006$.

With definitions (\ref{VVH},\ref{int},\ref{FF}) we calculate the
couplings and show results in Table 1. Note, that additional
interactions (\ref{VVH}) were already considered {\it e.g.} in
works~\cite{IAnd,VVH}.
\newpage
 Table 1. Coupling constants
$G_{VV'}\,GeV^{-2}$ of effective interactions $H V V'$ in dependence
on value of $\lambda$. All coupling values are multiplied by $10^7$.

\begin{center}
\begin{tabular}{||c|c|c|c|c||}
\hline $\lambda $  & $G_{WW}$
&$G_{ZZ}$
 & $G_{Z\gamma}$ & $G_{\gamma\gamma}$
 \\
\hline
$0.01$ & $ 3.10 $ & $4.40$ & $4.71$& $1.26$
\\
\hline
$0.006$ & $2.20$ & $3.10$ &  $3.32$& $0.89$
\\
\hline
$0.003$ & $1.33$ & $1.86$ & $2.00$& $0.54$
\\
\hline
$0 $ & $ 0 $ & 0  & 0& 0\\
\hline
$-0.003$ & $-3.51$ & $-4.83$ & $-5.17$&
$-2.58$\\
\hline
$-0.006 $ & $-6.54$ & $-9.01$  & $-9.65$& $-2.58$\\
\hline
$-0.01$ & $-10.3$ & $-14.2$ & $-15.2$&
$-4.08$\\
\hline \hline
\end{tabular}
            \end{center}

Then we calculate cross sections of pair weak boson production
accompanied by the Higgs. In doing this we apply the CompHEP
package~\cite{CompHEP}. In subsequent Tables we show results for
values of couplings in~(\ref{VVH}) and cross sections of processes
$p+p \to W^+W^- H+X$ ($\sigma(+-)$), $p+p \to W^+ Z H+X$
($\sigma(+0)$) and $p+p \to W^-Z H+X$ ($\sigma(-0)$). Results are
shown in dependence on value of $\lambda$ in admissible
interval~(\ref{CMSlam}).
\\
\\

Table 2. Production LO cross sections of $V V' H$  for
$\sqrt{s}=8\,TeV$ at the LHC.

\begin{center}
\begin{tabular}{||c|c|c|c||}
\hline $\lambda$  & $\sigma(+-)\,fb$
&$\sigma(+0) fb$
 & $\sigma(-0) fb$
 \\
\hline
$-0.01$ & $5.18$  & 1.21 & 0.499
\\
\hline
$ -0.006$ & $ 4.62 $ & $1.09$ & 0.45
\\
\hline
$-0.003$ & $ 4.22 $ & $1.02$ &  0.43
\\
\hline
$ 0$ & $ 3.86 $ & $0.98$ &  0.42
\\
\hline
$ 0.003 $ & $3.75 $ & $0.98$  & 0.42\\
\hline
$0.006$ & $ 3.69 $ & $0.98$ &  0.42
\\
\hline
$0.01$ & $ 3.62 $ & $0.98$ &  0.42
\\

\hline \hline
\end{tabular}
            \end{center}

 Table 3. Production LO cross sections of $V V' H$  for
$\sqrt{s}=13\,TeV$ at the LHC.

\begin{center}
\begin{tabular}{||c|c|c|c||}
\hline $\lambda$  & $\sigma(+-)\,fb$
&$\sigma(+0) fb$
 & $\sigma(-0) fb$
 \\
\hline
$-0.01$ & $21.16$  & 3.60 & 1.47
\\
\hline
$ -0.006$ & $ 17.06 $ & $2.71$ & 1.20
\\
\hline
$-0.003$ & $ 14.36 $ & $2.27$ &  1.07
\\
\hline
$ 0$ & $ 11.90 $ & $2.08$ &  1.00
\\
\hline
$ 0.003 $ & $11.14 $ & $2.09$  & 1.00\\
\hline
$0.006$ & $ 10.70 $ & $2.12$ &  1.02
\\
\hline
$0.01$ & $ 10.35 $ & $2.19$ &  1.03
\\

\hline \hline
\end{tabular}
            \end{center}
From Table 3 we see, that for $\sqrt{s}=13\,TeV$ with negative
$\lambda$ the effect is noticable, especially for process $p+p\to W^+
W^- H+X$, and {\it e.g} for $\lambda=-0.01$ the cross-section is
almost two times more than the SM one. However the cross-section
itself is presumably not sufficiently high for a productive study of
the effect.

Let us note also, that $V V' H$ the additional interaction with $\lambda
\neq 0$ might give effect for  VBF Higgs production. However
calculations show, that with couplings from Table 1 effects even for
$\sqrt{s} = 13\, TeV$ are insignificant, as well as effects for
branching ratios of the Higgs decays.

We have also calculated effects of the interaction for an associated single
electro-weak boson production with the Higgs. While cross sections of processes
are significant (few hundreds of {\it fb}), contributions of the additional
interaction do not exceed few {\it per cent}. For example, for
process $p\,p\to W^+\,H+X$ at $\sqrt{s}=13\,TeV$ with $|\lambda|=0.01$ ratio
$\mu=1.033$.

Possible manifestations of vertices~(\ref{VVH}) were studied in
decays $H
 \to W^+W^-,\,H \to Z\,Z$~\cite{VVH} . Results of this work give
limitations, which definitely do not contradict values for couplings
presented in Table 1.

In the next section we consider additional contributions of
interactions~(\ref{VVH}) to interaction of the Higgs with quarks,
especially with the heavy ones, which can lead to essentially more
significant effects at the LHC.

\section{Additional top and bottom quarks interaction with the Higgs}
\label{sec3} We use vertices~(\ref{VW}) to define additional
contribution for quarks interactions with Higgs. For the beginning we
shall be interested in interactions of the most heavy top quarks.
\begin{figure}
\includegraphics[width=7.5cm]{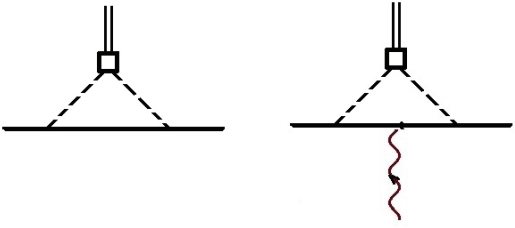}
\caption{Diagram representation of non-perturbative contribution
to $\bar t t H$ and $\bar t t H G$.
Double lines represent Higgs boson,
dotted lines represent electro-weak bosons.
Thick lines represents the $t$ quark, the wavy line corresponds to a gluon. Non-perturbative part
of $V V H$ vertex is defined in the previous section and simple points
correspond to usual gauge interaction.}
\label{fig:ttH}
\end{figure}
Taking into account these vertices we calculate loop diagrams
presented in Fig. 2 to obtain the following expression for $\bar t t
H$ vertex, which corresponds to the first diagram in
Fig.\ref{fig:ttH}
\begin{eqnarray}
& &V_{\bar t t H} = -\frac{g}{2 M_W}\,\bar t\,\bigl( M_t+9 \cos\theta_W M_Z M_W
\times\nonumber\\
& & G_{WW} I_1\,
(\hat p_1-\hat p_2)(1+\gamma_5)\bigr)\,t H;\; \hat a=a_{\mu}\gamma^{\mu};
\label{ttH}
\end{eqnarray}
 where $G_{WW}$ is already defined in~(\ref{VW}) and calculated in
Table 1, $p_1$ and $p_2$ are respectively outcoming momenta of $t$
and $\bar t$ quarks. Integral $I_1$ is defined in~(\ref{int}). For
calculation of the integral we here use the same form-factor
$F(t)$~(\ref{FF}).

Due to QCD gauge invariance we have to take into account also vertex
for fourfold interaction involving also a gluon: $\bar t t H
G_{\mu}$, which actually corresponds to the second diagram in
Fig.\ref{fig:ttH}
\begin{equation}
V_{\bar t t H G}=9 g g_s \cos\theta_W M_Z G_{WW} I_1\,\bar t \hat G
(1+\gamma_5) t H;\label{ttgH}
\end{equation}
where $g_s$ is the QCD gauge coupling constant and of course usual
structure of the QCD is used. Then we perform calculations for cross
sections of process $p+p \to \bar t\,t\,H+X$ for two energies of the
LHC: $\sqrt{s}=8\,TeV$ and $\sqrt{s}=13\,TeV$

Let us define for the same values of $\sqrt{s}$ ratios of
cross-sections with nonzero $\lambda$ in admissible
interval~(\ref{CMSlam}) and its SM value for $\lambda=0$
\begin{equation}
{\Large\mu}_{\sqrt{s}}\,=\,\frac{\sigma_{\lambda}(pp\to\bar t t
H)}{\sigma_{0}(pp\to\bar t t H)};\label{rat}
\end{equation}
where $\sigma_{0}$ is actually SM value for the cross section.
Results of calculations with application of CompHEP
package~\cite{CompHEP} are shown in Table 4.
\\

Table 4. Production LO cross sections $fb$ of $\bar t t H$  for
$\sqrt{s}=8\,TeV,\,13\,TeV$ and ratio ${\Large\mu}_{\sqrt{s}}$
(\ref{rat}) at the LHC.
\begin{center}
\begin{tabular}{||c|c|c|c|c||}
\hline $\lambda$  & $\sigma(13\,TeV)$
&$\sigma(8\,TeV)$
 & ${\Large\mu}_{13}$ & ${\large\mu}_8 $
 \\
\hline
$-0.01$ & $1628.2$  & $460.6$ & 3.14& 3.15
\\
\hline $ -0.006$ & $ 1212.9 $ & $342.6$ & 2.34 & 2.34
\\
\hline
$-0.003$ & $ 853.3 $ & $241.0$ &  1.65 & 1.65
\\
\hline
$ 0$ & $ 517.8 $ & $146.1$ &  1.00 & 1.00
\\
\hline
$ 0.003 $ & $401.4 $ & $113.6$  & 0.78 & 0.78\\
\hline
$0.006$ & $ 348.7 $ & $98.4$ &  0.67 & 0.67
\\
\hline
$0.01$ & $ 304.4 $ & $ 85.9$ &  0.59 & 0.59
\\

\hline \hline
\end{tabular}
            \end{center}
Values for cross-sections are calculated with the current value for
the strong coupling~\cite{pdg}
\begin{equation}
\alpha_s(M_Z)\,=\,0.1181\pm0.0011.\label{AS}
\end{equation}
The uncertainty in~(\ref{AS}) means 2\% accuracy for calculated
ross-sections. With taking into account of other sources of
uncertainties we estimate overall accuracy to be around 10\%.

The combination of the ATLAS and the CMS data, collected with
$\sqrt{s}=7\,TeV$ and $8\,TeV$, gives the following experimental
result for ratio ${\large\mu}_8$~\cite{AT+CMS}
\begin{equation}
{\large\mu}_8 \,=\,2.3^{+0.7}_{-0.6}\,.\label{Rexp}
\end{equation}
Due to significant uncertainties, result~(\ref{Rexp}) does not mean
convincing deviation from the SM value. With numbers from Table 4 we
have from~(\ref{Rexp}) the following prediction
\begin{equation}
\lambda\,=\,-\,0.0057^{+0.0028}_{-0.0039}\,.\label{lamres}
\end{equation}
The result is safely inside experimental limitation~(\ref{CMSlam}).
Let us note, that estimates for cut-off energy scale~(\ref{Lambda})
also evidently do not contradict LHC results~\cite{CMSlam}.
Of course we have in~(\ref{lamres}) again only two standard deviations effect,
which undoubtedly needs further studies. We see that values of
${\large\mu}$, practically, do not depend on $\sqrt{s}$ but with
$\sqrt{s}=13\,TeV$ cross sections are more than three times as much
as those for conditions of result~(\ref{Rexp}). One might hope to
check the predictions in forthcoming experimental studies at the LHC
with increased statistics. Emphasize, that in case of this study
would give result $\lambda \neq 0$, we would come to the fundamental
conclusion of non-perturbative effects in the electro-weak
interaction to be necessarily present. Let us note, that recent NLO
and NNLL SM calculations of $\bar t t H$ production cross section at
13 TeV are presented in works~\cite{br1, br2, ttHN}.

Let us consider also associative production of the Higgs with $\bar
b\,b$ pairs. Unlike the $t$ quark pairs case, for which the
experimental studies were performed and have given results, {\it e.
g.}~(\ref{Rexp}), there were no dedicated studies. However,
interactions~(\ref{ttH},\ref{ttgH}) in our consideration also exist
for other quarks. All the difference is connected only with value of
the quark mass in~(\ref{ttH}). In particular, it is advisable to
consider also process of $b$ quark associative pair production with
Higgs
\begin{equation}
p+p\to \bar b b H\,+\,X.\label{bbH}
\end{equation}
Results of calculations are shown in Table 5.
\\

Table 5. Production LO cross sections $fb$ of $\bar b b H$  for
$\sqrt{s}=8\,TeV,\,13\,TeV$ and ratio ${\Large\mu}_{b{\sqrt{s}}}$
 at the LHC.
\begin{center}
\begin{tabular}{||c|c|c|c|c||}
\hline $\lambda$  & $\sigma(13\,TeV)$
&$\sigma(8\,TeV)$
 & ${\Large\mu}_{b{13}}$ & ${\large\mu}_{b{8}} $
 \\
\hline
$-0.01$ & $1612.5$  & $572.7$ & 2.93 & 2.77
\\
\hline $ -0.006$ & $ 1204.4 $ & $433.6$ & 2.19 & 2.10
\\
\hline
$-0.003$ & $ 903.0 $ & $329.1$ &  1.64 & 1.59
\\
\hline
$ 0$ & $ 550.8 $ & $206.7$ &  1.00 & 1.00
\\
\hline
$ 0.003 $ & $446.0 $ & $151.8$  & 0.81 & 0.82\\
\hline
$0.006$ & $ 396.4 $ & $151.8$ &  0.72 & 0.73
\\
\hline
$0.01$ & $ 354.6 $ & $ 137.1$ &  0.64 & 0.66
\\

\hline \hline
\end{tabular}
            \end{center}
Effects for the b-quarks are of the same order of magnitude as for
the $t$-quark pairs. Unlike the $t$-quark case we have no datum for a
comparison. As a matter of fact, analogous results are valid for
light quarks $u, d, c, s$ as well.
\section{Conclusion}
The problem of an existence of non-perturbative contributions in the
electro-weak interaction is without doubt a fundamental one.
Anomalous three-boson interaction~(\ref{3boson}) provides the crucial
test for this problem. We have shown above, that there are promising
processes
\begin{equation}
p+p \to \bar t t H +X \quad p+p \to \bar b b H +X; \label{ttbb}
\end{equation}
for investigation of the problem at the LHC, and we can hope, that
future results for these processes at $\sqrt{s}=13\,TeV$ will confirm
the existence of non-perturbative effects in the electro-weak
interaction. Let us note, that we have studied how results
for processes under the study depend on different cuts. It comes out,
that for main process $p+p \to \bar t t H$ cuts  $M(\bar t t)>M_0,
M(t H)>M'_0, p_T(H)>p_{T0}$ {\it etc} lead, of course, to decreasing of cross
sections, but, practically,  do not change ratios $\mu$ for data.
Thus introduction of cuts is to be defined by conditions of experiments,
in particular by background considerations. 

The important problem is, if predictions of the present work could in
any way contradict the present knowledge. We have already mentioned,
that contributions of the additional interactions to branching ratios
of the Higgs are negligible.  The effects in process $p\,+p \to \bar
q q H+X$, where we have to take into account all six flavors of
quarks would lead to an additional contribution to the total Higgs
production cross section. For example, for $\lambda=-\,0.006$, which
actually is quite close to the central value in
estimate~(\ref{lamres}) we have from Tables 4, 5 the following
additional contributions $\Delta\,\sigma$ to the total cross section
of the Higgs production
\begin{eqnarray}
\Delta\,\sigma(8 TeV)= 1.36\,pb;\quad\sigma(8 TeV)= 22.3\,pb;\nonumber\\
\Delta\,\sigma(13 TeV)= 4.03\,pb;\;\sigma(13 TeV)= 50.6\,pb;\label{Delta}
\end{eqnarray}
where we also show the SM values for the total
cross-sections~\cite{pdg}. These additional contributions lead to a
change in the global signal strength, which currently
reads~\cite{pdg}
\begin{equation}
\mu=1.09\pm 0.07\pm0.04\pm0.03\pm0.07;\label{mu}
\end{equation}
where two last errors are connected with uncertainties in the
theoretical estimates. We easily see, that additional
contributions~(\ref{Delta}) give the following changes for
theoretical predictions for effective $\mu$ instead of unity
\begin{equation}
\mu\,(8 TeV)= 1.061\,;\quad\mu\,(13 TeV)= 1.080 .\label{Delta8,13}
\end{equation}
The results evidently do not contradict to value~(\ref{mu}), which is
based mostly on data collected with $\sqrt{s}=7$ and $8\,TeV$.

In case of an existence of triple interaction~(\ref{3boson}), {\it
e.g.} in processes~(\ref{ttbb}), being confirmed, extensive studies
of other possible non-perturbative effects will be desirable.

For example, effects in top pair production in association with an
electro-weak boson $W^\pm, Z$ were discussed in work~\cite{ptep}
under assumption of wouldbe existence of four-fold electro-weak
bosons effective interaction~\cite{BB1, BB2}.

\section{Acknowledgment}
The work is supported in part by grant NSh-7989.2016.2 of the
President of Russian Federation.

\end{document}